# SiliFuzz: Fuzzing CPUs by proxy

Kostya Serebryany
Google
kcc@google.com

Maxim Lifantsev
Google
maxim@google.com

Konstantin Shtoyk
Google
kostik@google.com

Doug Kwan
Google
dougkwan@google.com

Peter Hochschild
Google
phoch@google.com

## Abstract

CPUs are becoming more complex with every generation, at both the logical and the physical levels. This potentially leads to more logic bugs and electrical defects in CPUs being overlooked during testing, which causes data corruption or other undesirable effects when these CPUs are used in production. These ever-present problems may also have simply become more evident as more CPUs are operated and monitored by large cloud providers.

If the RTL ("source code") of a CPU were available, we could apply *greybox fuzzing* to the CPU model almost as we do to any other software [1]. However our targets are general purpose x86_64 CPUs produced by third parties, where we do not have the RTL design, so in our case CPU implementations are opaque. Moreover, we are more interested in electrical *defects* as opposed to logic *bugs*.

We present SiliFuzz, a work-in-progress system that finds CPU defects by fuzzing software proxies, like CPU simulators or disassemblers, and then executing the accumulated test inputs (known as the *corpus*) on actual CPUs on a large scale. The major difference between this work and traditional software fuzzing is that a software bug fixed once will be fixed for all installations of the software, while for CPU defects we have to test every individual core repeatedly over its lifetime due to wear and tear. In this paper we also analyze four groups of CPU defects that SiliFuzz has uncovered and describe patterns shared by other SiliFuzz findings.

## 1 Introduction

When discussing misbehaving CPUs we distinguish between *logic bugs* and *electrical defects*. A **logic bug** is an invalid CPU behavior inherent to a particular CPU microarchitecture or stepping, e.g. [2, 3, 4]. An **electrical defect** is an invalid CPU behavior that happens only on one or several chips. Often, but not always, an electrical defect affects only one of the physical cores on the chip. A defect may be present in a CPU from day one (if it was missed by the vendor during testing) or it may be a result of physical wear-out (e.g., circuit aging). For the purposes of this paper the distinction between day-one defects and wear-out is not important.

The term **"Silent Data Corruption" (SDC)** has been coined [5, 6, 7, 8] to represent CPU defects (and bugs) that do not cause an immediate observable failure, but instead silently lead to incorrect computation. This phenomenon has been known for a very long time, and is directly related to concepts like fault-secure operation, data integrity checking, and silent errors. In many production environments SDCs are more dangerous than crashes, because most server software ecosystems are built with tolerance to crashes, but not with tolerance to unobserved data corruption. Not all CPU defects induce SDCs, but many do.

In this paper we present SiliFuzz, a system for finding CPU defects. It uses coverage-guided fuzzing of software CPU simulators and disassemblers to generate a representative set of tests for CPUs (known as the *corpus*). We call this approach "fuzzing by proxy." Our underlying assumption is that if a test set provides good code coverage for the proxy, these tests are also likely to trigger interesting CPU behavior, and potentially uncover defects. We treat our test targets, CPUs, as opaque—we do not make any assumptions about the underlying structure of a CPU, nor do we use any vendor-internal debugging capabilities. Similar approaches have been used before for other targets. For example, a test set generated by fuzzing the Linux kernel USB subsystem was used to find bugs in the Windows kernel USB subsystem [9]. Our approach is also a kind of differential testing [10], since we cross-validate different CPU cores on many machines.

We do not claim novelty in the academic sense—all components of our system, such as fuzzing hardware as software and fuzzing an opaque system via proxy, have been known before [1, 9]. We also do not compare our system to others that are similar, as we do not have access to any such systems. Furthermore, the system we present here is early work in progress and it needs more work to reach maturity. Nonetheless, SiliFuzz is already producing results and we feel it is important to share our results with a wider audience.

The rest of the paper is organized as follows. We start with an overview of SiliFuzz in Section 2, provide a brief evaluation in Section 3, cover the related literature in Section 4, and discuss the future direction of work in Section 5 before concluding. Appendices A-D describe four groups of defects found in CPUs from two major x86_64 vendors.

## 2 SiliFuzz Overview

In this section we describe the SiliFuzz system in its current work-in-progress form. It consists of the following components:

- **Preparation:**
  - *Snapshot (§2.2):* A format for representing and serializing short, self-contained programs and their reference output(s).
  - *Player (§2.3):* A program for executing snapshots one-by-one, on a CPU or a simulator.
  - *Snapshot generation (§2.4):* A pipeline for generating sets of snapshots, with the goal of covering as many corner cases as possible.
- **Execution (§2.5):** A system to check every core of every machine in the fleet.
- **Feedback loop (§2.6):** A feedback loop between known bad machines and SiliFuzz.

SiliFuzz works completely in user space and thus does not directly test kernel-specific instructions.

### 2.1 Fuzzing and Coverage

First, we need to describe the basics of software fuzzing. Fuzzing is a technique of testing a target (an application, or an API) with a large number of test inputs generated on the fly. The goal is to make these inputs as interesting and as diverse as possible in order to trigger corner cases. In other words, fuzzing aims to maximize the combined code coverage. "Code coverage" may have different meanings, from basic block coverage to bounded path coverage. In this paper we use "coverage" as a synonym of software control flow edge coverage. The two most popular approaches to software fuzzing are *generation-based fuzzing* and *coverage-guided mutation-based fuzzing*.

Generation-based fuzzing (e.g. for software: [11, 12], or for hardware: [13, 14]) randomly creates an infinite number of test inputs, typically based on knowledge of the input grammar. Each new input is produced from scratch, not using any kind of feedback from the system under test. This is frequently called **blackbox fuzzing**.

Coverage-guided mutation-based fuzzing starts from a *seed corpus* of test inputs and applies small random mutations and crossovers to random elements of the corpus. The mutants are executed by the target application and the coverage feedback is collected. If a given mutant triggers previously unseen coverage it is added to the corpus. Guided fuzzing is typically performed by general-purpose fuzzing engines not developed for one specific target, such as libFuzzer [15], AFL [16], or Honggfuzz [17]. This is frequently called **greybox fuzzing**.

A variation of coverage-guided fuzzing, often called *structure-aware fuzzing*, requires that the mutations are applied to inputs in a way that preserves validity and syntactical correctness. Syzkaller [18] is one such fuzzing engine. libFuzzer allows users to define their own mutators, turning it into a structure-aware fuzzing engine.

Many fuzzing engines, blackbox and greybox, use *dictionaries*—user-defined byte sequences that are likely to trigger interesting behaviors if inserted into the test inputs.

## 2.2 Snapshot

A SiliFuzz *snapshot* describes instructions and data necessary for executing some relatively short sequence of binary code. Our typical snapshot contains less than 100 bytes of code and runs in microseconds, but it can be arbitrarily large. Valid snapshots do not contain syscalls or other sources of nondeterminism. They are executed in a single thread.

A snapshot consists of the following parts:

- Initial state of all user readable CPU registers, including the program counter.
- Memory mappings with their read/write/execute permissions and the corresponding data bytes. Note that snapshots expected to cause page faults require that particular pages **not** be mapped.
- Expected end state(s) of the execution. If a snapshot executes differently on different CPU microarchitectures it will have multiple end states. If the execution causes a CPU exception this is captured as the corresponding signal number and signal address.
- Additional metadata describing how this particular snapshot was produced (e.g. whether it was captured from a program, produced by combining several snapshots, etc).

## 2.3 Player

The *player* is a program that can replay a snapshot, capture the outcome and determine if the outcome matches one of the expected end states. We pin the execution to a particular core. A mismatch indicates a potential defect in the CPU core that executes the snapshot. The player is separated into two processes referred to as the *driver* and the *harness*. The driver is a stateful execution orchestrator. It performs operations like transforming snapshots into a series of commands and processing execution results. The harness performs commands issued by the driver including running the actual code. Having two processes provides the necessary isolation between the driver and potentially crashing code of the snapshot being played.

Consider a simple snapshot consisting of a single NOP instruction placed at the memory address 0x10000000. It will cause the following sequence of commands to be generated by the driver and executed by the harness:

1. MapMemory       { start = 0x10000000,
                     num_bytes = 4096 }
2. WriteMemory     { start = 0x10000000,
                     data = "\x90\xCC\x00...",
                     num_bytes = 4096 }
3. ProtectMemory   { start = 0x10000000,
                     perms = r-x }
4. ExecuteSnapshot { registers = < ...,
                     rip = 0x10000000 > }
5. ChecksumMemory  { start = 0x10000000,
                     num_bytes = 4096 }

The \xCC byte is an x86 trap instruction marking the end of the snapshot. The rest of the memory page at 0x10000000 is filled with zeroes to guarantee deterministic execution.

## 2.4 Snapshot Generation

SiliFuzz leverages open-source x86_64 CPU emulators and disassemblers as proxies to generate snapshots with the help of libFuzzer. The intuition behind this approach is that if a particular input sequence triggers interesting behavior (new coverage) in the proxy system then the same input may also trigger interesting behavior in a real CPU. We also generate random instruction sequences with ifuzz [19].

### 2.4.1 XED

XED [20] is an x86_64 instruction encoder/decoder. It can decode up to 15 bytes of data into one x86_64 instruction at a time, and report the number of decoded bytes. As such, it makes it a good target for fuzzing with libFuzzer (see Listing 1).

Listing 1: XED libFuzzer integration

```c
int LLVMFuzzerTestOneInput(const uint8_t *data, size_t data_len) {
  xed_decoded_inst_t xedd;
  xed_tables_init();
  xed_decoded_inst_zero(&xedd);
  xed_decoded_inst_set_mode(&xedd, XED_MACHINE_MODE_LONG_64,
                            XED_ADDRESS_WIDTH_64b);
  xed_decode(&xedd, data, data_len);
  return 0;
}
```

The result of fuzzing XED is a corpus of byte sequences which represent valid and non-valid x86_64 instructions. The corpus maximizes the coverage of XED itself, and thus likely triggers interesting behaviors in the CPU instruction decoder or other components. Appendix B is a demonstration of such interesting behavior.

### 2.4.2 ifuzz

The ifuzz library [19] is a part of the Syzkaller project. It provides both generation and mutation facilities for x86_64 instructions based on a model of the x86_64 ISA derived from XED. When used in generation mode, ifuzz only creates random *valid* instruction sequences. Since ifuzz is not guided by coverage, it essentially produces an infinite number of outputs with no way to choose a good subset. We use a sample of ifuzz-generated instruction sequences in two ways: first, we create snapshots from them to test on real CPUs and second, we use them as a dictionary for fuzzing Unicorn.

### 2.4.3 Unicorn

Unicorn [21] is a lightweight multi-platform, multi-architecture CPU emulator framework based on QEMU. We applied libFuzzer to Unicorn in a setup similar to the fuzz target available in the Unicorn source tree [22] to generate instruction sequences. In essence, this fuzz target maps the fuzzer-generated data at a fixed address and starts simulated execution from the first byte.

libFuzzer is oblivious to x86_64 instruction encoding rules and the nature of how the result is processed. However, during the fuzzing process most of the inputs added to the corpus are those that trigger new coverage in Unicorn and are thus mostly valid instruction sequences.

### 2.4.4 Snapshot Fixing and End State Recording

Fuzzing via XED, ifuzz, and Unicorn does not produce snapshots directly—rather, it produces instruction sequences. These sequences may contain syscalls and other sources of nondeterminism, and perform invalid memory access or contain illegal instructions. They also lack a well-defined end state. We apply a number of techniques to convert these instruction sequences into proper snapshots. These can be broken down into several stages:

1. **Adding and removing memory mappings.** We intercept memory page faults and map a limited number of additional pages accessed by a snapshot. The mapped pages are currently filled with zeroes.
2. **Nondeterminism elimination.** We eliminate all system calls by tracing all snapshots. We then verify that every snapshot successfully and deterministically replays multiple times on many machines. Each snapshot is limited to 3 seconds of CPU time by default.
3. **End state determination.** Finally, each snapshot is run on all supported Google production platforms (microarchitectures) and the per-platform end states are captured. Currently, we only keep snapshots that have identical end states on all microarchitectures as additional protection against hard-to-detect sources of nondeterminism.

We do not eliminate illegal instructions, primarily because there is no way to understand whether an instruction is illegal without executing it on all CPU generations. Appendix B demonstrates another reason for not removing illegal instructions.

## 2.5 Execution at Scale

### 2.5.1 SiliFuzz Checker

For running SiliFuzz at scale, we created a program, SiliFuzz *checker*, that efficiently runs multiple snapshots. This checker is used in background testing of production machines, described below. The checker reads snapshots from a corpus residing on a disk, materializes snapshots in memory and sends them to a pool of snapshot players, one for each core being tested. Our current corpus consists of very short snapshots. Reading a snapshot from a file system and re-materializing it in memory is therefore expensive compared to the execution time of the snapshot. To amortize the cost of reading snapshots, the checker groups snapshots in batches and reuses snapshots in a batch multiple times. Multiple batches are also executed concurrently to hide I/O delays. In our current setting, we batch 50 random snapshots together and create a random list of 1000 snapshot executions out of the batch using a uniform random distribution. These numbers are picked empirically to reduce the reading cost by a factor of 20. Execution of snapshots on the list are distributed evenly across all available cores using multiple queues.

### 2.5.2 Background Testing for Production Machines

To efficiently screen the fleet, SiliFuzz needs to spend a non-trivial amount of time on every core of every machine. Machines in Google's data centers are continuously screened for SDC defects with a collection of tests, including SiliFuzz. This is done while machines are in use in production. To minimize the performance impact of this screening, only a small number of machines in a cluster are being tested at any given time. On a machine being tested, one of the testing tools, for example SiliFuzz, is picked randomly based on a probability distribution customized for the machine's architecture. When SiliFuzz is chosen, job scheduling software on the machine reserves 4 cores (2 hyperthreaded pairs) for the test and allows SiliFuzz to run for up to 2 minutes. The core count, time limit, and testing frequency are chosen for all tests to keep screening overhead below a certain percentage of machine utilization. The job scheduling software also ensures the 4-core window eventually covers all cores on the machine. After SiliFuzz finishes, results are logged into a machine database for analysis.

## 2.6 Using Known Bad Machines for Feedback

Machines that are believed to be defective are taken out of production and set aside for detailed diagnostics. These machines are put into a quarantine pool so that we can run a number of tests on them. Testing usually takes weeks for a machine since some failures are dependent on frequency, voltage and temperature and we need to vary these parameters to run our tests with different combinations. The quarantine pool is a valuable tool for isolating defective machines in our fleet but it is also an important tool for SiliFuzz development. It allows us to:

- Compare SiliFuzz to other SDC detection tools and measure its relative performance and effectiveness. The population of the pool is not constant—machines enter and leave the pool over time, so we have to perform the comparison regularly.
- Avoid SiliFuzz regressions. In addition to comparing SiliFuzz against other tools, we can also compare different versions of SiliFuzz. This is especially useful for regression testing.
- Enrich the corpus via known bad machines. Our hypothesis is that a snapshot that uncovers one defect on one machine has a good chance to uncover some defect on another machine. Our data has an early indication supporting this hypothesis. So, we are preserving such snapshots in our corpus, regardless of their proxy coverage.
- Establish defect patterns. The quarantine pool serves as a source from which defect patterns can be extracted. Some failures recorded there are analyzed by a human to see if there is any commonality.

The pool includes machines where the Jump Conditional Code (JCC) [2] issue can be reproduced. A specially crafted snapshot can trigger the problem with high probability allowing us to continuously and reliably test the SiliFuzz infrastructure.

## 3 Evaluation

Our current corpus contains snapshots generated by fuzzing Unicorn and XED with libFuzzer and a number of random snapshots produced by ifuzz—about 500,000 in total. The corpus has been deployed to a large portion of the Google production fleet and detected multiple faulty machines. We only consider a machine defective if we are able to reproduce the same problem on the same core(s) over multiple days. **In about 70% of the cases a single pair of logical cores sharing the same physical core is affected.** In most other cases only a single logical core is affected. In just a handful of cases we observe many different cores failing. Some snapshots have detected defects on more than one machine, although those defects are not necessarily equivalent. We were able to group some of the defects detected by different snapshots based on similar behavior and likely similar causes.

Some machines flagged by SiliFuzz have been independently confirmed as defective by other tests. Others have been implicated in production incidents. Finally, about 45% of SiliFuzz findings are unique and have not been previously identified by any other tool or automation available to us. The data we collected fleetwide shows a wide distribution of time-to-failure (CPU time consumed by SiliFuzz on a given machine before detecting the first defect) with some machines failing almost immediately and others taking days to detect a single defect.

We have observed that our approach of fuzzing by proxy finds more defects compared to purely random snapshots produced by ifuzz. Preliminary data collected from the quarantine pool shows that from the machines identified by SiliFuzz, about 40% were uniquely identifiable by snapshots produced from Unicorn and XED. About 20% were similarly identified only by ifuzz-produced snapshots. The remaining 40% of the machines were detectable by both types of snapshots. This suggests that the two types of snapshots are largely complementary. Our insight is that CPU defects are common enough that any sufficiently non-trivial fuzzing will uncover some defects when applied at large scale. The twice as many defects found by fuzzing trivial proxies compared to using a sophisticated generation-based fuzzer suggests that fuzzing by proxy is worth further investigation, especially searching for better proxies.

Analyzing CPU defects is a labor-intensive task that is often hindered by poor reproducibility and complicated processes around getting access to specific production machines. In Appendices A-D we describe four groups of defects that we analyzed in detail. The first three affect more than one machine each. The fourth defect falls into a pattern that we observe on multiple machines, but the precise behavior is unique. Defect A was uncovered by a Unicorn-produced snapshot, Defect B was uncovered by a XED-produced snapshot, and Defects C & D were independently uncovered by ifuzz- and Unicorn-produced snapshots. We cannot guarantee this attribution with certainty, because early in the project we did cross-pollination between the three corpora. Going forward, we will need to more carefully track attribution of defects in the presence of cross-pollination.

We have detected many other defects that we did not analyze in detail, however we noticed some common patterns:

- Wrong results for integer, floating point and vector instructions. Often the result has just one incorrect bit.
- Spurious `SIGFPE`, `SIGSEGV`, or `SIGILL`.
- Missing floating point exception.
- No `SIGILL` where one should be raised according to the specification.
- Sticky flags, i.e. certain bits in `EFLAGS` or similar status registers are always set despite the lack of the underlying condition.
- Undershooting `REP(NE)`-prefixed string instructions.
- Floating point data pointer register (RDP) value mismatch.

We do not know whether these common patterns are representative of all CPU defects, or are simply easy to find using our technique. However we observe a correlation between types of defects we find most frequently and the parts of Unicorn that are implemented as C code (floating point, vector instructions, etc) as opposed to being just-in-time compiled, and our method provides better coverage feedback for the C code, than for the JIT-compiled code.

## 4 Related Work

The SDC problem with electrical defects in microprocessors was known from at least the 1960s. It became more widely known at least two decades ago [23, 5]. Today, for large service providers owning multiple data centers, SDC-inducing defects are a very real problem. Facebook discusses their SDC issue in [6] while Google provides a high level overview of SDC in [7].

Miller, Fredriksen, and So [24] first introduced fuzz testing. Since then, there has been much progress in this area. Readers can refer to [25] for a recent survey of fuzzing techniques. Our work draws on some previous work. In essence, SiliFuzz is a form of differential testing [10], which uses identical test inputs to discover differences between test subjects. It also heavily relies on advances in coverage-guided mutation-based (or "evolutionary") fuzzing [26].

There has been some previous work in fuzzing CPU hardware. Sandsifter [27] uses depth-first-search-based fuzzing and page fault analysis to exhaustively determine x86 instruction lengths on CPUs from various vendors. This is done to detect undocumented instructions and hardware bugs. Sandsifter focuses on instruction encoding as opposed to validating the functionality. Easdon [28] describes a way of detecting undocumented instructions only accessible to an operating system's kernel. UISFuzz [29] is another hardware instruction set fuzzer for undocumented instructions similar to Sandsifter. It uses instruction format information to speed up search. Like Sandsifter, it focuses on instruction encoding.

Generation-based fuzzing has been applied to hardware for several decades [14, 13]. However, in the past, it typically did not utilize coverage feedback to guide mutations. More recently, Trippel et al. [1] have translated an RTL design into a software model and applied off-the-shelf coverage-guided software fuzzing engines to the model. Our work uses a real CPU as the fuzz target, for which we do not have the original RTL design. A similar approach to ours was adopted by Martignoni et al. [30] but in the opposite direction. They used a real CPU as the golden reference to perform differential testing to find bugs in CPU emulators.

While pre- and post-silicon manufacturing tests are performed by CPU vendors before CPUs are delivered to customers [31, 32, 13, 14, 33], SiliFuzz aims to find defects in CPUs *already deployed* in production systems. Crucially, unlike system-level testing performed by CPU vendors, SiliFuzz has to work without the vendors' expertise, tools, and knowledge and instead leverage a large fleet of deployed CPUs at scale—which often is infeasible in pre-production verification.

## 5 Future Work

We are just starting our work on finding CPU defects. Future work will be concentrated along four axes: scale, speed, automation, and quality, with quality being the most challenging.

**Scale**

We need to run on as many cores for as much time as possible. This is expensive, and we are looking for ways to reduce the cost.

**Speed**

The player needs to be further optimized to execute more snapshots per second, and reduce syscalls and I/O operations. Slow-executing snapshots need to be removed without sacrificing the combined proxy code coverage.

### Automation

Detected defects need to be automatically analyzed, deduplicated and reported. Defective machines need to be automatically evicted from the fleet without false positives. CPU vendors need to be provided with actionable reports, reproducers, and often the hardware itself.

### Quality

Quality of the corpus is a function of two values: coverage it provides, and time needed to execute it (or, simply, its size). If the corpus is too large, it becomes too expensive to execute all of it on all cores. If the coverage is low, we will not find many defects. There are many ways we can increase the coverage while maintaining a tolerable size.

**Mutation strategies.** We currently mutate the instruction sequences as simple byte sequences. This is interesting from the point of view of stressing the CPU instruction decoders (or maximizing the coverage of XED), but produces too many invalid instruction sequences. They are mostly not added to the corpus since they generate little new coverage, but they slow down the corpus generation due to the high volume of such snapshots under the current fuzzing strategies. We will need to add structure-aware mutations that produce only valid x86_64 sequences. Specialized snapshot mutation strategies are also expected to have a strong effect (see e.g. "register scrambling" from [34]). Cross-pollinating snapshots generated by different proxies is likely to have a positive effect, and we already do it in a limited way via dictionaries.

**Corpus distillation,** a process of choosing the smallest corpus subset while maintaining coverage. Individual corpora coming from Unicorn and XED are already distilled, but their combination may have redundancies. We currently do not distill the ifuzz corpus at all. As we start using more sources of snapshots (e.g. harvesting snapshots from real programs) we will need an efficient distillation strategy, via better coverage metrics and better proxies.

**Better coverage metrics.** Code coverage, while the best we have for software testing, is not a very strong metric of corpus quality. Even when a piece of software has 100% control flow edge coverage, it may still have bugs not detectable by the test corpus. Complete code coverage of the software proxy in our approach guarantees even less when testing the hardware. We may be able to develop or reuse better metrics specifically for fuzzing CPUs. We will continue to rely on software proxies, but ideally we would also collect coverage metrics directly from hardware. However, we are likely to remain constrained by the CPUs being mostly opaque to us.

**Better proxies.** XED only allows fuzzing the decoder-like functionality of CPUs. Unicorn is a functional CPU simulator and thus does not give any coverage feedback for the microarchitectural behavior. We will need to find ways to fuzz microarchitectural models, e.g. based on gem5 [35] or Simics [36].

**Quality of the player.** The player currently executes multiple snapshots in the same process (harness) with non-trivial set up code in between. We expect that minimizing this set up code will improve our chances to find defects since every snapshot execution will then inherit random microarchitectural state from previously executed snapshots. We can improve further by running multiple non-overlapping snapshots in parallel threads, such that we stress cache coherency and similar subsystems.

**Multi-state.** Currently we discard a considerable number of snapshots (up to 20%) because they contain multiple end states. These can be caused by nondeterminism in these snapshots or by meaningful microarchitectural differences. Being able to differentiate the two (e.g. by statically analyzing the disassembled code of each snapshot) will further improve the quality of our snapshot corpus.

**Bugs vs. defects.** SiliFuzz may potentially be used for finding CPU bugs. Unlike electrical defects, logic bugs can be detected on a single CPU, or even on a CPU model, before the actual CPU even exists. For example, our system can easily reproduce the JCC bug [2] and thus, *theoretically*, can find it too. However, in order to reliably detect bugs we need to allow snapshots with multiple end states and distinguish valid differences between end states (e.g. one CPU family not supporting a given instruction) from differences caused by a bug.

# 6 Conclusion

In this paper we presented our work-in-progress system for finding CPU defects. We fuzz proxies, such as software CPU simulators and disassemblers, using traditional software fuzzing techniques, and then use the accumulated corpus to cross-check millions of CPU cores. We have detected a large number of defects, analyzed four of them in detail, and analyzed common patterns among the others.

We expect this and similar technologies to be in widespread use in the coming years since CPU defects are here to stay.

# Acknowledgements

We thank our colleagues Mike Fuller, Eric Liu, Marco Elver, Dmitry Vyukov, Andrew G. Morgan, Chris Leary, Mike Gunter, Subhasish Mitra, and Eric A. Schmidt Jr. for the help with preparing this paper.

# Appendices

## A  F2XM1 Defect

The first defect we found with SiliFuzz is related to the F2XM1 x87 instruction, which computes $2^x - 1$ for a given value of $x$. We observed on several CPUs that this instruction *sometimes* returns incorrect results. Listing 2 shows a code snippet from one of the snapshots detecting this issue.

Listing 2: F2XM1 Defect
```
fldln2 // load natural log(2) on
       // stack top.
f2xm1  // compute 2^x - 1 using
       // stack top value.
```

Upon investigation, it turned out that only the F2XM1 instruction matters. It does not matter if the input comes from an FLDLN2 instruction or loaded from memory. On a given machine, we observed that failures only appear *occasionally* on the defective core. The incorrect results seem to be constant for a given physical core but vary between affected cores. The precise underlying failing mechanism is still unknown to us. We have also confirmed the fault with a vendor-supplied diagnostic tool, which reports F2XM1-related issues on those cores we found.

This defect looks similar to the one described by Dixit et al. [6], but we do not know if it is the same.

## B  Overshoot of an Illegal Instruction

The second defect found by SiliFuzz is a class of issues related to illegal instructions. x86_64 supports numerous instruction prefixes that can be applied only to some instructions, and not to others. In normal circumstances, a prefix followed by an incompatible instruction would always raise an illegal instruction fault (#UD). We observed some machines sometimes overshoot past these ill-formed instructions without raising a #UD fault.

These snapshots originated from fuzzing the x86 disassembler, which explains why they contain illegal instructions. While this is technically a defect, as the CPUs involved did not behave according to specifications, the problem does not affect normal operation of our fleet because typical production code does not contain illegal instructions. It is still possible that the same defect has another, more harmful, manifestation that we did not observe.

## C  FCOS Miscomputes

The third defect we found involves FCOS and other transcendental functions. Listing 3 shows an instruction sequence that sometimes produces incorrect results on one of the machines screened by SiliFuzz. The sequence comes from a hand written reproducer inspired by failing snapshots.

Listing 3: FCOS Miscomputes
```
// Prologue
fldln2
fldln2
fprem1
// Payload
fld1
fcos
```

The expected value of the above should be 0.5403023059 (or `0x3fe14a280fb5068c` in binary representation) but the actual value is 0.5403023043 (binary `0x3fe14a280edd9be7`). The difference is less than 0.0000003%. The erroneous value is consistent for both logical cores sharing the physical core.

One interesting aspect of this defect is that **there seems to be some hidden state or memory**. The prologue part of the instruction sequence seems to put the core in a broken state and then the payload starts miscomputing. **The prologue can be run in the same process as the payload or in a different process.** The CPU recovers from the broken state itself after an unknown period of time. We have found ways to trigger miscomputation even without the prologue. Simply running the payload in a loop with different values on ST(0) starts miscomputing after some number of iterations.

An attempt to map the values that miscompute is presented in Figure 1. Our current theory is that bit 23 (i.e. `1<<23`) in the binary encoding of the input argument to FCOS triggers the problem: when the bit is set the computation works as expected but when it is reset the instruction almost universally miscomputes on affected cores.

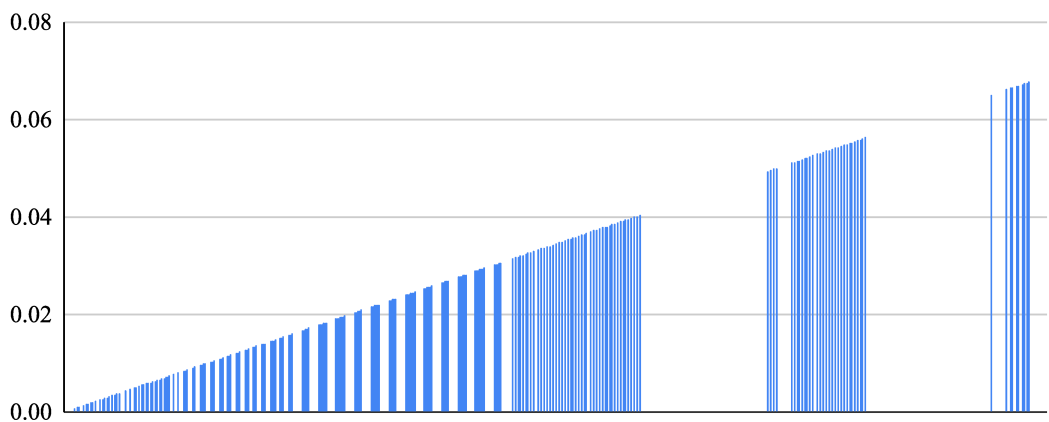

Figure 1: Miscomputing FCOS. *Axis X is the argument to COS, Y is X if miscomputed, 0 otherwise.*

## D  Missing x87 Data Pointer Update

We observed a defect in which one of the x87 error registers is not consistently updated. The legacy x87 floating-point state contains a register for the last data address used by a non-control x87 instruction. In the example sequence seen in Listing 4, the FISTTPLL instruction should set the data pointer to the store address. On the machine on which we found this defect, the data pointer is instead sometimes set to zero. This missing update only happens for the data pointer but not other error registers and this symptom is only found on a particular core pair of the machine. We believe this is a genuine hardware defect.

Listing 4: Missing x87 Data Pointer Update

```
cmp     %esi,%edx
rcr     $0xf3,%ecx
fldpi
fisttpll (%rdi) # should set x87 dp
adc     (%rax),%eax
```